\documentclass[prb,twocolumn,showpacs,preprintnumbers,amsmath,amssymb]{revtex4}

\usepackage{graphicx} 
\usepackage{dcolumn}
\usepackage[xdvi]{epsfig}
\usepackage[usenames]{color}
\usepackage{amsfonts}
\usepackage{subfigure}
\usepackage{txfonts}
\usepackage{dcolumn}
\usepackage[latin1]{inputenc}

\definecolor{MyLightMagenta}{rgb}{0.1,0.2,0.8}
\let\oldmarginpar\marginpar
\renewcommand\marginpar[1]{\-\oldmarginpar[\raggedleft\large{\color{MyLightMagenta} \vspace{-12pt} #1}]{\raggedright\large{\color{MyLightMagenta} \vspace{-12pt} #1}
}}

\begin{document}

\title{Study of the volume and spin collapse in orthoferrite LuFeO$_3$ using LDA+U}
\author{Donat J. Adams}
\email{donat.adams@cea.fr}
\author{Bernard Amadon}%
\affiliation{CEA, DAM, DIF, F 91297 Arpajon, France}
\date{\today}

\begin{abstract}
Rare earth (R) orthoferrites RFeO$_3$ exhibit large volume transitions associated with a spin collapse. We present here \emph{ab initio} calculations on LuFeO$_3$. We show that taking into account the strong correlation among the Fe-3\emph{d} electrons is necessary. Indeed, with the LDA+U method in the Projector Augmented Wave (PAW), we are able to describe the isostructural phase transition at 50~GPa, as well as a volume discontinuity of 6.0\% at the transition and the considerable reduction of the magnetic moment on the Fe ions. We further investigate the effect of the variation of $U$ and $J$ and find a linear dependence of the transition pressure on these parameters. We give an interpretation for the non-intuitive effect of $J$.  This emphasizes the need for a correct determination of these parameters especially when the LDA+U is applied to systems (e.g in geophysical investigations) where the transition pressure is \emph{a priori} unknown.
\end{abstract}

\maketitle

\section{Introduction \label{sec:int}}

The magnetic spin collapse under pressure in transition metal oxides has attracted great interest in the past few years, not only because of the geophysical implications\cite{kantor2005, speziale2005, lin2005} and fundamental questions on the origin of the transition, but also because experiments and structural calculations have become feasible in the pressure regions where the spin transition takes place (see for example Refs.~\onlinecite{yoo2005, adams2005, kasinathan2006, kunes2008}). These high pressure regions could be reached because of the development of the diamond anvil cell (DAC)\cite{fiquet2001} on the experimental side. The development of electronic structure codes allowing for structural relaxations facilitates the computational treatment. In particular, plane wave methods combined with the Projector Augmented Wave (PAW) framework\cite{blochl1994, kresse1999, holzwarth1997, torrent2008} allow to treat atoms throughout regions, where the ionic radii, the ionic positions, the nominal valence as well as the crystal structure might heavily vary. 
However, the standard treatment using the Density Functional Theory (DFT)\cite{hohenberg1964, kohn1965} in the local spin density approach (LSDA) is erroneously cumbered by the so-called self-interaction energy and  more generally, by the wrong description of interactions of electrons inside localized orbitals (e.g $3d$).

The limitation of DFT methods has provoked the development of new theoretical methods, such as the LDA+U and LDA+DMFT methods\cite{anisimov1991, anisimov1997, georges1996, kotliar2006} (from a combination of the DFT in the Local Density Approximation (LDA) and a Hubbard Hamiltonian), the self-interaction corrected LSDA (SIC-LSDA),\cite{temmerman1998} or the hybrid functional method.\cite{heyd2004} 

Recently spin and volume collapse isostructural transitions under pressure have been observed in orthoferrites\cite{xu2001, rozenberg2005} with X-ray-diffraction methods and M\"ossbauer spectroscopy. LuFeO$_3$ is an ideal material to test the agreement between experimental and computational methods. The reasons for this are manifold: First, accurate experimental data exist up to pressures of 125~GPa.\cite{rozenberg2005} Second, the transition is well defined contrary to some other orthoferrites (e.g. PrFeO$_3$). Third, this compound has a simple magnetic structure,\cite{white1969} because of the complete $f$-shell of lutetium: correlation effects inside the $f$-shell can thus be neglected and the focus can be put on the correct description of the iron atom. 
Fourth, the distortion of the perovskite structure in LuFeO$_3$ is strong which clearly defines the geometrical structure of the crystal and allows to neglect thermal effects on the crystal structure (in other perovskite materials\cite{zhong1996} such as SrTiO$_3$ and BaTiO$_3$ this distortion is smaller and leads to a sequence of thermal phase transitions). This simplifies the theoretical treatment and allows straight forward comparison with experimental results. 
Fifth, the stoichiometry of LuFeO$_3$ perovskite is well defined in contrast to other materials (e.g. FeO) where it is difficult to obtain pure samples.\cite{ono2007, kantor2003} 

As a large number of transition-metal compounds are insulators, the existence of correlated metals raises many theoretical questions which have been tackled recently (see e.g. Refs.~\onlinecite{pavarini2004, pavarini2005, raychaudhury2007}.) However, orthoferrites are Mott insulators with a large gap\cite{xu2001, rozenberg2005, singh2008} and the rare earth ions retain an important atomic behavior.

In general, the high spin (HS) to low spin (LS) transition in these systems is linked to the considerable volume collapse of the transition metal ion.\cite{anderson1989} It leads to the violation of atomic first Hund's rule because of the enhanced crystal field and thus to the considerable reduction or even complete vanishing of the magnetic moment. The success of recent calculation using LDA+U and LDA+DMFT to describe volume and moment collapse in simple oxides such as MnO\cite{kasinathan2006, kunes2008} and CoO\cite{wdowik2008} emphasizes that this transition is clearly linked to the existence of strong interactions.

The electronic structure of LuFeO$_3$ has been studied within the PBE formalism by \citet{xing2007} and \citet{iglesias2005} Though, the volume collapse transition was not studied in these works. In PBE \cite{iglesias2005, xing2007}, the AFM magnetic structure is found correctly but no band gap \cite{iglesias2005} is found or it is small (0.46~eV\cite{xing2007}). Indeed the R-FeO$_3$ orthoferrites are known for their large optical gap.\cite{xu2001, pasternak2002} In Ref.~\onlinecite{xing2007} atomic relaxations were applied whereas in Ref.~\onlinecite{iglesias2005} the calculations were performed in the ideal cubic geometry, which could explain the different findings. Our work supports the notion of Ref.~\onlinecite{iglesias2005}: The combination of a standard treatment of the electronic exchange-correlation energy (LDA here, GGA in Ref.~\onlinecite{iglesias2005}) with the cubic crystal structure results in a metallic state. Recently, \citet{singh2008} have performed calculations using the LDA+U formalism: much larger gaps are obtained. Unfortunately, the value of Coulomb interaction $U$ is not given.

In our study we thus focus on the phase transition, and use the LDA+U approximation which describes well Mott insulators such as LuFeO$_3$. We show that the method is indeed able to describe the volume collapse associated to the spin transition upon pressure. We study how $U$ and $J$ contribute to the stabilization of the two phases.

\begin{figure}
\begin{center}
\includegraphics[width=0.45\textwidth]{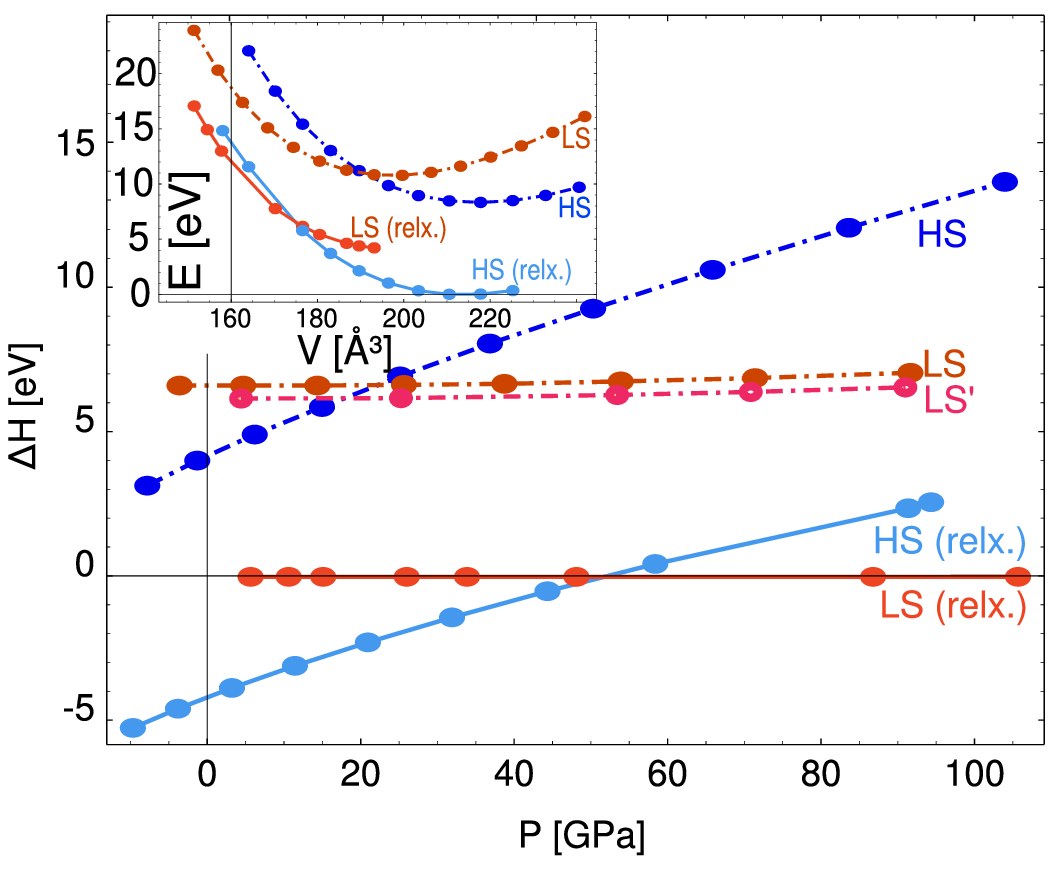}
\caption{ \label{pic:res:dh_lufeo3_comp} The enthalpy of the HS, the LS and the LS' phase of LuFeO$_3$ in the pressure range of 0--100~GPa using relaxed (relx) and unrelaxed LDA+U calculations. The enthalpy $H^{LS}(p)$ obtained from the fit of the BM3 EOS was used as a reference. Further details in the text. Inset: the energy of the HS and LS phase at the corresponding volumes.
  }
\end{center}
\end{figure}

\section{ Methodology \label{sec:meth} }

In this section we gathered the description of the PAW datasets, the computational details related to LuFeO$_3$, an analysis of the atomic occupation matrices for the HS and LS phases in the LDA+U method and a discussion of the thermodynamic potential at pressure, the enthalpy.
 
\subsection{ PAW atomic datasets \label{sec:meth:pot} }

PAW atomic datasets are generated using ATOMPAW.\cite{holzwarth2007, tackett2001} For the LDA+U PAW calculations, semi-core states of Lu and Fe are treated in the valence. Valence states for Lu, Fe and O thus include 5\emph{s}5\emph{p}5\emph{d}4\emph{f}6\emph{s}, 3\emph{s}3\emph{p}4\emph{s}4\emph{p}3\emph{d} and 2\emph{s}2\emph{p} states, respectively.
PAW radii are 2.52 a.u., 2.01 a.u. and 1.11 a.u., respectively. These values of the radii have been chosen in order
to avoid the overlap of PAW augmentation regions at high pressure.\\

The Lu, Fe and O atomic data were tested for the oxygen molecule (O$_2$, box size 10 $a_0 \simeq 5.3$~\AA, energy cutoff 16~Ha$\simeq$ 545~eV), bcc ferromagnetic iron (k-points: $11\times 11 \times 11$, 56 points in total, energy cutoff 20 Ha $\simeq$ 545~eV), rhombohedral iron oxide and hexagonal Lu metal (k-point: $19\times19\times13$, 585 points in total, energy cutoff 20 Ha $\simeq$ 545~eV). Equilibrium properties are compared to values in the literature in Tab.~\ref{tab:pottes} which validates the PAW atomic data. In the case of lutetium metal, calculations of Ref.~\onlinecite{pickard2000} are carried out with the ultrasoft-pseudopotential method, which could explain the discrepancy with the more precise PAW method, especially for a system which contains strongly localized $f$ electrons (mainly contained in the augmentation region).

In order to test the validity of the PAW datasets we performed LDA calculations on rhombohedral FeO. A four atomic rhombohedral (\emph{R}-3\emph{m}, space group 166) unit cell was chosen in order to accommodate antiferromagnetic ordering.\cite{cococcioni2005} We use 30 k-points for the k-point sampling and a 16~Ha plane wave energy cutoff. Thus $a_0$ is converged to 0.001~\AA{ }and $B_0$ to 7~GPa. The resulting EOS parameters of the AFM phase are given in Tab.~\ref{tab:pottes}. 

\begin{table}
\caption{
\label{tab:pottes} 
Summary of the PAW atomic data. The values given represent our calculated quantities. The data of Ref.~\onlinecite{pickard2000} are obtained from ultrasoft pseudopotential LDA calculations.}
\begin{tabular}{l l r@{.}l r@{.}l r r@{.}l}
\hline
Molecule/crystal   & quantity & \multicolumn{2}{c}{Our work} & \multicolumn{3}{c}{Literature}  & \multicolumn{2}{c}{error \%}  \\
\hline
 O$_2$   &  d$_0$~(\AA{ }) & 1&216 & 1&21 &  Ref.~\onlinecite{serena1993}  & 0&4\\
 Fe ferromagnetic   &  a$_0$~(\AA{ }) & 2&76 & 2&76 &  Ref.~\onlinecite{garcia2004} & 0& \\
    &  $\mu (\mu_B)$ & 1&99 & 2&08 & & 0&06  \\ 
    Lu metal hcp  &  $V_0$~(\AA$^3)$ & 52&24& 53&373&  Ref.~\onlinecite{pickard2000}& 2&\\
    &  $c/a$ & 0&641 & 0&640 & &  0&1\\ 
    FeO  &  $a_0$~(\AA$^3)$ & 4&185& 4&179&  Ref.~\onlinecite{alfredsson2004}& 0&12\\
    &  $B_0$ [GPa] & 242& & 237& & &  2&\\
\hline
\end{tabular}
\end{table}

\subsection{ Calculation setup \label{sec:meth:gen} }

\subsubsection{Computational scheme}
Calculations are performed using the ABINIT package,\cite{gonze2002} within the PAW \cite{torrent2008} framework. The electron-electron interaction is treated using the LDA and the LDA+U.\cite{anisimov1991, anisimov1997, liechtenstein1995, czyzyk1994, bengone2000, amadon2008} The unit cell of the ideal perovskite structure is cubic and contains 5 atoms. In order to establish the GdFeO$_3$ orthorhombic distortion observed by \citet{rozenberg2005} a unit cell has to be considered with lattice vectors $e_1+e_3$, $2e_2$ and $e_1-e_3$, where $e_i$ are the lattice vectors of the ideal cubic perovskite. This unit cell contains 20 atoms and allows for antiferromagnetic ordering of the Fe atoms. This unit cell was used for our unrelaxed antiferromagnetic calculations while the symmetry elements were reduced to those of space group \emph{Pbnm}. For the unrelaxed structure, we found the antiferromagnetic (AFM) configuration to be lower in energy than the ferromagnetic (FM) configuration  by 1.458~eV in good agreement with experiment where the AFM is found for the HS and the LS phases.\cite{rozenberg2005} \footnote{These calculations were performed using $U=$4.3~eV and $J=$0. The volume was 217.822 \AA$^3$ which corresponds to a pressure of $\simeq$1.5~GPa.} 
This AFM symmetry was therefore imposed during the calculation which on the other hand reduced the degrees of freedom. On the Lu positions we find no magnetic moment and therefore no magnetic ordering.

Convergence of LuFeO$_3$ was reached on a $6\times4\times6$ Monkhorst sampling grid (36 k-points) and with an energy cutoff of 16 Ha ($\approx435$~eV). Energy differences of the HS and the LS phase were converged to 10 meV. The pressure is converged to 0.08 GPa and enthalpies to 50 meV. Moreover, transition pressures are converged to 0.04 GPa. The double counting energy in the fully localized limit (atomic limit)

\begin{equation} \label{eqn:setup_1}
E_{\mathrm{dc}}[{n^\sigma}]=U/2N(N-1)-J/2[N^\uparrow(N^\uparrow-1)+N^\downarrow(N^\downarrow-1)]
\end{equation}

was used throughout this work, where $N^{\sigma}= \mathrm{tr} \; n_{ij}^{\sigma}$ and $N=  N^{\uparrow}+N^{\downarrow}$.\cite{liechtenstein1995, czyzyk1994, anisimov1993} The parameter $U$ representing the Coulomb repulsion of the Fe-3\emph{d} orbitals was chosen 4.3~eV as it was determined by \citet{cococcioni2005} for the oxide FeO. A higher value could be used because of the more contracted orbitals
in Fe$^{3+}$ with respect to Fe$^{2+}$.\cite{solovyev1996} However,
in other works\cite{mosey2008} a slightly lower value was determined (3.7~eV). The values are thus difficult to compare because they depend on the orbital basis set. For the sake of simplicity and in order to compare with the Ref. \onlinecite{cococcioni2005}, the exchange parameters $J$ was set to 0.\cite{mazin1997} Spin-orbit coupling is neglected in these calculations. We assume that the orbital magnetic moment -- although not negligible -- is mainly quenched by the crystal field. 

\subsubsection{Determination of the ground state}

In the ionic limit, the charge of lutetium, oxygen, and iron would be  3+, 2- and 3+. This implies a formal occupancy of 5 electrons for the \emph{d}-orbital sub-shell. In a cubic lattice, the ground state would thus consist of the filling of 3 $t_{2g}$ and 2 $e_g$ orbitals (Hund's Rule requires maximal spin polarization), while in the LS state electrons fill 5 $t_{2g}$ states. In the atomic limit --- corresponding to the complete filling of orbitals  and without fluctuations ---, and thus in LDA+U, the symmetry would therefore be broken in the LS state.\cite{larson2007} Calculations for both phases have thus been carried out by imposing the \emph{Pbnm} symmetry to the electronic states even for the undistorted structure. The experimentally observed crystal distortion appears not to be the consequence of the electron distribution in the LS phase alone, as the distortion is also present in the high spin phase. It is probably more due to the geometric redistribution of space between the ions of different radii as described in section \ref{sec:res:geom}.

While for the HS configuration only one electronic arrangement is possible, the LS can be implemented in several ways because in the \emph{Pbnm} symmetry the degeneracy of all \emph{d}-orbitals is lifted. In order to enhance the convergence of the specific spin state the electron-electron interaction potential was fixed in the Hamiltonian according to a given occupancy matrix $n_{\rm ij}^\sigma$ (see \ref{sec:meth:gen})\footnote{See also Ref.~\onlinecite{jomard2008} for a similar computational scheme} for the $d$ correlated subspace during the 30 first steps of the energy minimization procedure. Then it was self-consistently optimized. We determined energies and orbital occupancies starting from one hundred different occupation matrices (for the correlated subspace)  at a volume of 156.92~\AA$^3$ ($\simeq90 $~GPa). These one hundred occupations matrices correspond to the number of possibilities to have 3 spin majority and 2 spin minority states in the $d$-shell.

The optimization of the electronic density ended in 3 possible electronic states which are all $t_{\rm 2g}^{\uparrow 3}t_{\rm 2g}^{\downarrow 2}$. These 3 electronic states corresponds to different possible coupling of the $t_{2g}$ orbitals. We compared the energies of the 3 electronic configurations at pressures from 0 to 90~GPa and found that the enthalpies differ by a constant, which is independent of the pressure. This indicates an identical EOS for the three states in the LS spin region (see also FIG.~\ref{pic:res:dh_lufeo3_comp}). Assuming only diagonal $t_{2g}$ density matrix as starting point, we find that the configuration $d_{xy}^1d_{yz}^2d_{xz}^2$ is the most stable one.\footnote{x,y,z axis are parallel to Fe-O bonds in the cubic unit cell.} It is  referred to as LS in the following. The use of starting density matrix containing $e_{g}$ occupations (among the one hundred used) has enabled us to study a larger set of solution. Among them, the configuration LS' has the lowest energy ($E_{\rm LS}-E_{\rm LS'}=0.52$~eV at a volume of 156.92~\AA$^3$). The configuration in the majority channel is $d_{xy}^1d_{yz}^1d_{xz}^1$. In the minority channel two $t_{2g}$ states are occupied and characterized by non-diagonal occupation matrix elements. It gives a transition pressure which is 4~GPa lower than the one of LS. The LS and the LS' structure converge towards a unique state upon structural relaxation. This is reflected in the energy as well as in the occupation matrix of the Fe-3\emph{d} electrons. The third LS state (LS$^*$) lies between LS' and LS, as far as the energy is concerned. In its minority channel, only one of the $t_{2g}$ states contributes to non-diagonal matrix elements in the occupation matrix.
\\
Atomic relaxations were performed using the Broyden-Fletcher-Goldfarb-Shanno minimization.\cite{schlegel1982} Enthalpies were converged to a precision of at least $0.2$~meV, energies to $10^{-3}$~meV, forces to 0.01 $eV/$\AA. 

\subsubsection{The thermodynamic potential at pressure}
The 3rd order Birch-Murnaghan (BM3)\cite{poirier2000,birch1947} equation of state
 
\begin{equation}
\begin{split}
 E(x)&=E_0+\frac{9B_0V_0}{16} [B_0'(x^{2/3}-1)^3 +(x^{2/3}-1)^2\cdot (6-4x^{2/3})] \\
 p(x)&=\frac{3B_0}{2} (x^{7/3}-x^{5/3})\cdot [1+\frac{3}{4}(B_0'-4)(x^{2/3}-1)] \\
 x&=\frac{V_0}{V}
\end{split}
\end{equation} 

was fitted to the energy-volume data, where $E_0$ is the ground state energy, $V_0$ the ground state volume, $B_0$ the bulk modulus and $B_0'$ its first volume derivative.

In density functional theory the correct density is the one which corresponds to the lowest energy. Two different phases can correspond to local minima of the energy surface $E(V)$, where $E$ is the total energy and $V$ the volume.

However, the free parameter in experiment is not the volume $V$  but the pressure $p$ applied to the sample. The corresponding thermodynamic potential is the enthalpy $H(p)=E(V(p))+p\,V(p)$. The enthalpy can be computed if the pressure $p$ is determined from the electronic density using the Hellman Feynman theorem as it is done in ABINIT.  The pressure at which the transition occurs is given by the equality of the enthalpies. The transition pressure can be equivalently given by the  well known construction of common tangents on $E(V)$ curves.

In practice, the discrete data set \{$p_i,\,H_i$\} of various phases can be compared to the continuous enthalpy $H_r(p)$ of an additional phase, which can be chosen to be the enthalpy of the fitted $E(V)$ curve with the BM3 expression. The enthalpy difference $\Delta H^{\mathrm{s}}=H^{\mathrm{s}}-H_r$  (where $\mathrm{s} \in \{\mathrm{HS,LS}\}$)
allows to study the effects that lead to phase transitions and leave away features which are common to all phases considered (FIG.~\ref{pic:res:dh_lufeo3_comp}). In particular, the transition pressure is given by the intersection of two curves.

\begin{figure}
\begin{center}
\subfigure[]{\includegraphics[width=0.45\textwidth]{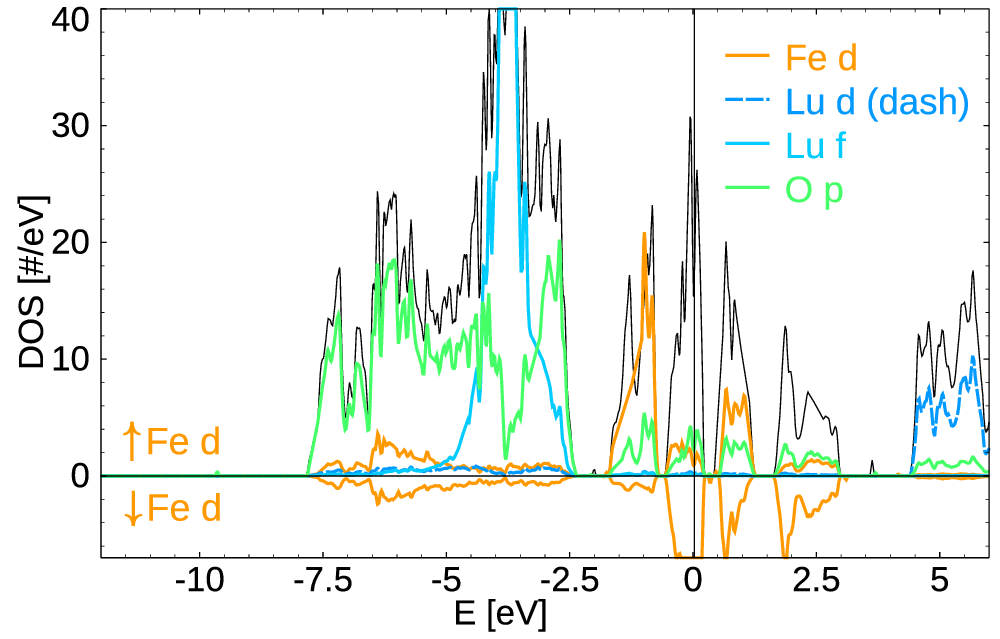} \label{pic:met:lda_r_GPa_ls_0_pardos}}
\subfigure[]{\includegraphics[width=0.45\textwidth]{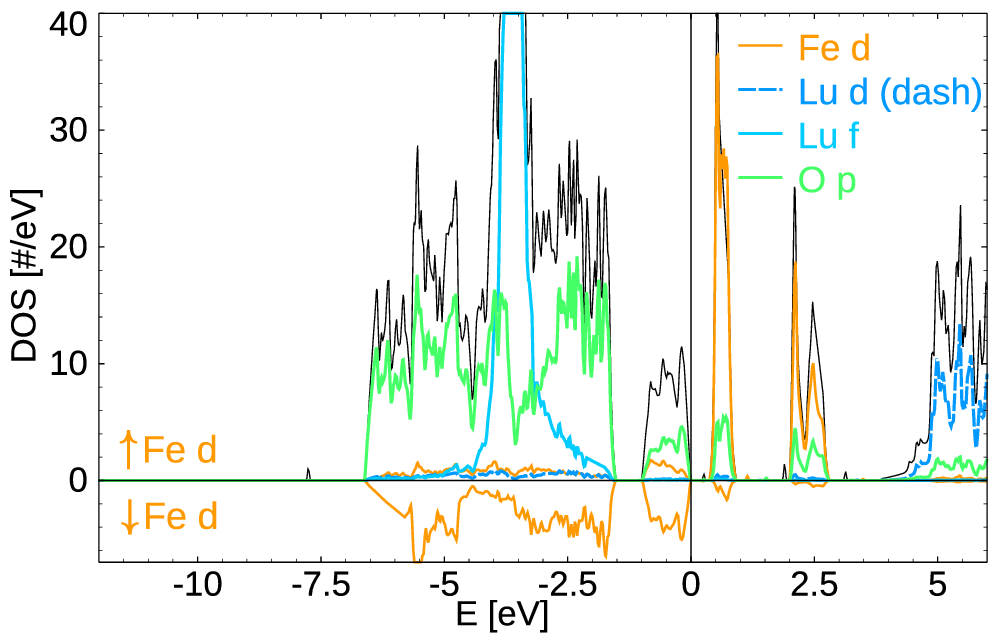} \label{pic:met:lda_r_GPa_hs_-10_pardos}}
\caption{
\label{pic:met:lda_dos}
Total density of states of from LDA calculations (for explanation of symbols see caption FIG. \ref{pic:met:ldau_dos}). 
\subref{pic:met:lda_r_GPa_ls_0_pardos} The relaxed LDA structure at a pressure of -3~GPa which corresponds to a volume of 197.458~\AA$^3$.
The gap vanishes and the local magnetic iron moments 1.1~$\mu_B$.
\subref{pic:met:lda_r_GPa_hs_-10_pardos} The relaxed LDA at a pressure of -12.3~GPa which corresponds to a volume of 224.882~\AA$^3$.
The gap is 0.7~eV and the local magnetic moments on Fe are 3.56~$\mu_B$. 
}
\end{center} \end{figure}

\begin{figure}
\begin{center}
\subfigure[]{\includegraphics[width=0.45\textwidth]{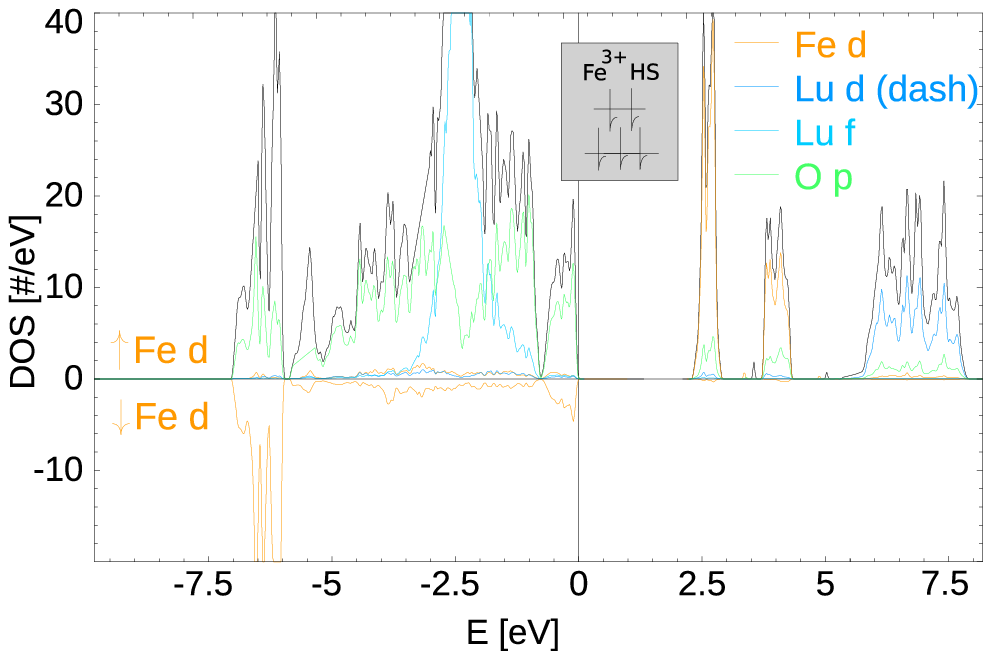} \label{pic:met:hs_GPa70_pardos}}
\subfigure[]{\includegraphics[width=0.45\textwidth]{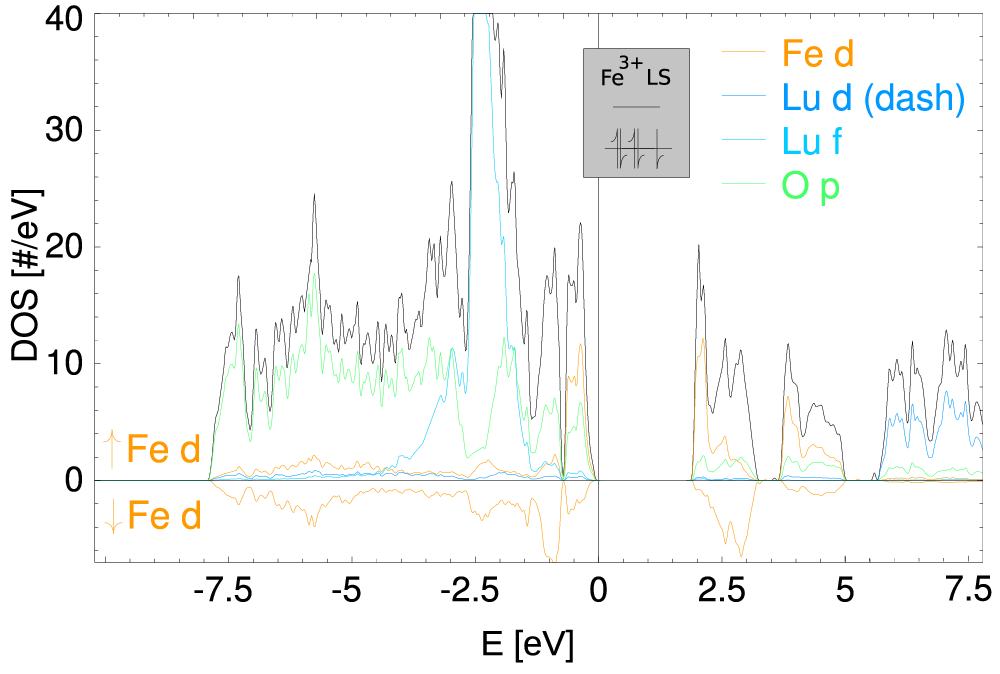} \label{pic:met:ls_GPa130_pardos}}
\caption{
\label{pic:met:ldau_dos}
Total density of states of LuFeO$_3$ (black) and projected densities on Lu, Fe and O from LDA+U calculations. For Fe the spin density is shown, positive values for the minority channel ($\uparrow$), negative values for the majority channel ($\downarrow$). Lu-\emph{f} electrons remain localized and hardly hybridize. For better visibility these peaks were cut.  
\subref{pic:met:hs_GPa70_pardos} The relaxed HS phase at a pressure of 3.2~GPa which corresponds to a volume of 210.54~\AA$^3$.
\subref{pic:met:ls_GPa130_pardos} The relaxed LS phase at a pressure of 48.1~GPa which corresponds to a volume of 170.21~\AA$^3$.
}
\end{center} \end{figure}

\section{Results \label{sec:res}}

This section presents the description of the volume-collapse transition in LDA+U.

\subsection{Electronic properties and densities of states \label{sec:meth:dos} }

We first have performed LDA calculations for the ideal cubic configuration and have found the resulting DOS to be metallic and in particular \emph{d}-states are at the Fermi level. Even starting from converged LDA+U ionic structures and the wave functions of the HS and the LS phase, respectively, the LDA fails: At small pressures (large volume) it gives a metallic DOS with a small magnetic moment of $ 1.1 \mu_B$ (FIG.~\ref{pic:met:lda_r_GPa_ls_0_pardos}). Only at an extremely large volume (highly negative pressure, FIG.~\ref{pic:met:lda_r_GPa_hs_-10_pardos}) a gap opens (0.7~eV) and a considerable magnetic moment results (3.56~$\mu_B$). This electronic configuration though is not stable at smaller volume (in particular not at the volume corresponding to ambient conditions). This emphasizes the incorrect description of correlations. Indeed, rare earth orthoferrites are know to be large gap insulators.\cite{xu2001, pasternak2002, singh2008} It is not possible to stabilize two different phases at positive pressure within the LDA.

As described above (section \ref{sec:meth:gen}) and emphasized before (e.g Ref.~\onlinecite{kasinathan2006}), the LDA+U method, which takes into account the strong correlations in the atomic limit, is able to describe these two phases. The density of states (DOS) of the HS and LS phase -- at a volume belonging to their individual stability range -- are shown on FIG.~\ref{pic:met:hs_GPa70_pardos} and \ref{pic:met:ls_GPa130_pardos}. The main effect of LDA+U is to stabilize the HS phase at positive pressures and to increase the gap inside the \emph{d}-orbitals. 
In the LS and the HS phases, the Fe-3\emph{d} orbitals in the valence band are strongly hybridized with the oxygen \emph{p} orbitals. The Lu-\emph{f} retain their atomic-like character while the Lu-\emph{d} --  though unoccupied -- hybridize with O-\emph{p} orbitals. The DOS scarcely changes with pressure, even at low pressures (0~GPa) where the LS phase is unstable. The HS phase (FIG.~\ref{pic:met:ls_GPa130_pardos}) is characterized by atomic-like peaks for all elements, especially Lu and Fe.

At 0~GPa ($U=4.3$~eV and $J=0$, no relaxation) the HS phase is insulating with a gap of 1.75~eV, and the LS with a gap of 1.16~eV. At 50~GPa the gaps are 1.10~eV and 0.93~eV, respectively. In fact, the gap size shows a perfect linear behavior in the range between 0 and 200~GPa. Relaxation increases the band gap at 0~GPa to 2.17~eV (HS) and 2.03~eV (LS). In Ref. \onlinecite{singh2008}, the value of the gap of the HS phase obtained in LDA+U is $\simeq8$~eV. The comparison with our results remains difficult because the parameters $U$ and $J$ were not given there. 

One of the major impacts of the parameters  $U$ and $J$ is their influence on the gap size. While $U$ always increases the gap, $J$ decreases it. The effect on the HS phase is half as strong as on the LS, where the gap increases by 0.45~eV when $U$ increases by 1~eV and decreases by 0.85~eV when $J$ increases by 1~eV.

The local spin moment on the Fe atoms is found to be 4.05~$\mu_B$ in the HS and 0.99~$\mu_B$ in the LS structure at 0~GPa, which decreases to 3.83~$\mu_B$ and 0.91~$\mu_B$ at 100~GPa, respectively (100~GPa corresponds to $V=$164.79~\AA{ } in the HS and $V=$155.00~\AA{ } in the LS phase). 

The discrepancy between the ionic ideal value of the spin moment and the actual value can be explained as follows: The hybridization of Fe-\emph{d} electrons with oxygen-\emph{p} electrons generates small occupancies on orbitals which are empty in the ionic picture. $e_g$ orbitals are more hybridized with oxygen-\emph{p} states: they are thus more filled. In the HS phase, 2 of the hybridized $e_g$ contribute to the minority spin and thus to the considerable reduction of the magnetic moment with respect to the ionic value (i.e. 5~$\mu_B$). In the LS phase, the hybridized $e_g$ orbitals appear in both spin channels and cancel each other. Only an almost empty $t_{2g}$ orbital reduces the local magnetic moment, which therefore remains close to the ionic value.

As will be detailed in the next section, at ambient pressure we describe an insulator-insulator transition for LuFeO$_3$. Recent LDA+U calculations on MnO give a similar conclusions, although the LS state obtained in this work seems to be less intuitive.\cite{kasinathan2006}
However we emphasize here the limitation of our work which does not contains fluctuations. These fluctuations could easily make the system metallic as has been shown recently in MnO.\cite{kunes2008} Moreover, recent experimental works on NdFeO$_3$\cite{gavriliuk2003} and BiFeO$_3$\cite{gavriliuk2007, gavriliuk2008} show that the transition is closer to an insulator-semiconductor or an insulator-metal transitions. Further optical experiment or LDA+DMFT calculations on LuFeO$_3$ could clarify this issue.

\subsection{The equation of state and the transition pressure} \label{sec:res:trans}

As emphasized before, the LDA is not able to describe the occurrence of two phases. Nevertheless, a continuous and linear decrease of the local magnetic moment from 1.05~$\mu_B$ to 0.55~$\mu_B$ is observed between 0 and 100~GPa (relaxed structure). 

LDA+U calculations for the two phases were first performed with the frozen ideal cubic configuration on the 20 atoms super-cell. The symmetry was reduced to the 8 elements of the corresponding \emph{Pbnm} space group in order to allow lifting of the degeneracy of the Fe-\emph{d} states. The energy and enthalpy versus volume curves are reproduced on Fig~\ref{pic:res:dh_lufeo3_comp}. One can see a phase transition between the LS and the HS phases. It occurs at 22~GPa. 

\begin{figure}
 \begin{center}
 \includegraphics[width=0.45\textwidth]{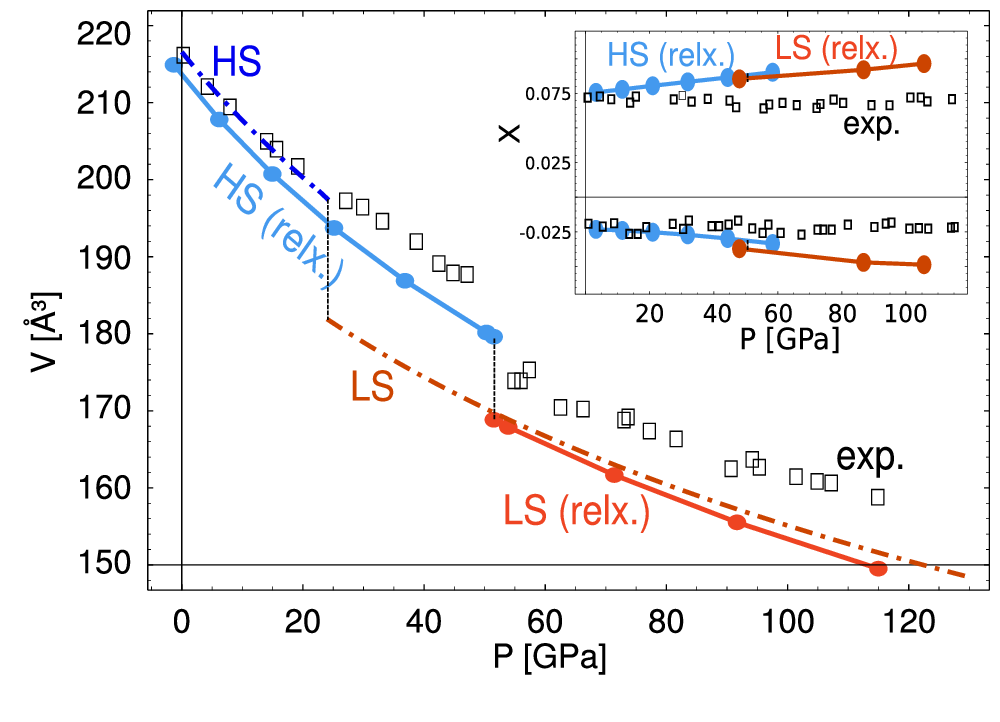}
 \caption{\label{pic:res:lufeo3_p_V_comp}
The volume of LuFeO$_3$ at pressures between 0 and 120~GPa. Bullets ($\bullet$): Calculated values using LDA+U with parameters $U=$~4.3 and $J=$~0~eV and performing structural relaxations. Dashed lines: Calculated values using the same parameters $U$ and $J$ without structural relaxation. The calculated HS-LS transition occurs at 21~GPa. Boxes ($\Box$): Experimental data from Ref.~\onlinecite{rozenberg2005}. Inset: Reduced atomic coordinates of Lu, bullets ($\bullet$) calculated, boxes ($\Box$) experimental data from Ref.~\onlinecite{rozenberg2005}. The symbols correspond to the X and Z coordinates, respectively.
  }
 \end{center}
\end{figure}

This pressure shifts to 51~GPa after structural relaxations, which compares well with the experimentally observed transition pressure of 50~GPa (FIG.~\ref{pic:res:dh_lufeo3_comp}).\cite{rozenberg2005} Note, that the HS-LS transition was not found in earlier works and the Fe magnetic moment was underestimated (3.6 $\mu_B$)\cite{xing2007}. The calculated collapse of the ground state volume at the transition amounts to 6.0\%, compared to the measured 5.5\%. Whereas in LDA the volume at zero pressure is underestimated by 12\%, the introduction of interactions localizes the electrons, and thus raises the volume: the volumes at zero pressure are then well reproduced (within 2.5\%, see Tab.~\ref{tab:eosparam}) for both the HS and LS phases. Bulk moduli are in the range of experiment but with an error of about 20\%. As a consequence, the difference between theoretical volume and experimental one increases to  6\% at 150~GPa (see Fig~\ref{pic:res:lufeo3_p_V_comp}). Besides, the discontinuity of the bulk modulus at the transition is qualitatively described (Tab.~\ref{tab:eosparam}): 31~GPa (this work) compared to 72~GPa in experiment\cite{rozenberg2005} (see also FIG.~\ref{pic:res:lufeo3_p_V_comp}).

Positional parameters are in good agreement with the experimental findings, moreover, we find no significant discontinuity of the positional parameters at the transition pressure as observed experimentally (FIG.~\ref{pic:res:lufeo3_p_V_comp} inset). 

We investigated the effects of the Hubbard part of the Hamiltonian on the transition pressure. We therefore computed energies (non-self-consistently) in the LDA frame from the densities obtained in LDA+U ($E_{\rm LDA}[n_{\rm LDA+U}]$) and observed that the HS-LS transition is maintained but shifted to much lower pressures. This shows that the LDA+U potential creates two different electronic densities which give different LDA energies. The $E_{\rm U}$ term in the energy additionally determines the difference of energy between the two configurations, LS and HS  (see also the discussion in section \ref{sec:res:trans_uj}).

Finally, we mention that we have neglected the entropic contribution in these calculations. In the iso-structural transition in cerium, the entropy appears to be essential to describe the transition.\cite{amadon2006} In LuFeO$_3$ we could expect the variation of entropy to be rather weak also because of the broken symmetry of $d$ states. However, we could miss fluctuations between configurations in the LS case. Furthermore, theoretical studies beyond LDA+U (e.g LDA+DMFT) as well as experimental studies of the transition as a function of temperature could help to understand these issues. 

\begin{table}
\caption{\label{tab:eosparam} Parameters of the fitted  3$^{\rm rd}$ order Birch Murnaghan EOS of the relaxed HS and LS phase of LuFeO$_3$ LDA+U ($U=4.3$~eV and $J=0$) . $E_{\rm 0,HS}-E_{\rm 0,LS}=$3.25~eV. The gap at $P=$0~GPa is also given.
}
\begin{ruledtabular}
\begin{tabular}{ l c  | c  c  c l}
& & \multicolumn{1}{l}{$V_0$ [\AA$^3$]} &   \multicolumn{1}{l}{$B_0'$} & $B_0$ [GPa] & Gap [eV]\\
\hline
 LDA&                &   195.06 &  4.19   & 236 & 0 \\
LDA+U & HS           &   213.59 &   3.80   & 214 & 2.17 \\
LDA+U & LS           &   197.41 &   3.77   & 245 & 2.03 \\
 Exp.\cite{rozenberg2005} & HS &   218.40     & \multicolumn{1}{c}{}     &  241  \\
 Exp.\cite{rozenberg2005} & LS &   199.40     &  \multicolumn{1}{c}{}    &  313 \\
\end{tabular}
\end{ruledtabular}
\end{table}

\subsection{The ground state structure}
\label{sec:res:geom}
Perovskites (ABX$_3$) are known to consist of mostly rigid BX$_6$ octahedra,
while the A cations are placed in the interstitial space between the octahedra. Tilting of the mostly rigid octahedra allows to decrease space assigned to the A cations and the unit cell volume. The tilting angle is sensitive to the ratio of the B-X bonding length and the ionic radius of A.\cite{woodward1997b,lufaso2001} According to \citet{glazer1972} and \citet{woodward1997a} perovskites can be classified according to three tilting angles and the phase (+/-) of successive octahedra along the tilting axis (see Ref. \onlinecite{glazer1972} p.~3386 for a sketch of the tilting system and Ref. \onlinecite{woodward1997a} p.~34 and p.~36 for further details on this concept). LuFeO$_3$ belongs to the space group \emph{Pbnm} and the tilting system is $a^+b^-b^-$. We find tilting angles of $a\simeq b\simeq13$~° in the HS phase at 3~GPa and $a\simeq 17$~° and $b\simeq 11$~° in the LS phase at 87~GPa. The tilting angles are only approximate, because a symmetry conserving distortion is superimposed to the tilting of the octahedra and in the distorted octahedra the determination of the tilting angle is not unique. While in the HS phase the octahedra are elongated by 1\% in the LS phase they are squeezed by 3\%. The relaxed atomic positions of the HS and the LS phase can be found in Tab.~\ref{tab:atpos}. The more important distortion of the octahedra in the LS spin phase could be attributed to the existence of a Jahn Teller effect in this phase.
\\

\begin{table}
\caption{\label{tab:atpos} Atomic Wyckoff positions of LuFeO$_3$ in the HS and the LS phase (space group \emph{Pbnm}, no. 62). The lattice parameters of the HS at 0~GPa are 5.156~\AA, 7.491~\AA{ }and 5.530~\AA{ }compared to the experimental values of \citet{marezio1970} of 5.213~\AA, 7.565~\AA{ }and 5.547~\AA{ }(maximal deviation for $a$ by -1.1\%, which shows that the chosen $U$ parameter cannot fully correct the known overbinding behaviour of LDA). The lattice parameters of the LS at 50~GPa are 4.617~\AA, 6.896~\AA{ }and 5.323~\AA. Compared to the data published by \citet{marezio1970} for the HS phase the average positional deviation is 0.013~\AA, and the biggest deviation is for Lu by 0.03~\AA. The unit cell volume is underestimated by 2.4\% in the calculations,\cite{marezio1970} while the experimental ratio of the unit cell vectors a:b:c of 0.689:0.733:1 is reproduced by 0.688:0.738:1 in the calculation.  
}
\begin{ruledtabular}
\begin{tabular}{ l | c l l l }
Atom & Wyckof position & x & y & z \\
\hline
 &\multicolumn{4}{c}{HS at 0~GPa, this work} \\
\hline
Lu  & 4c & 0.0233 & 0.     & 0.0758 \\
Fe  & 4a & 0.     & 0.     & 0.      \\
O 1 & 4c & 0.3783 & 1/2    & 0.0439 \\
O 2 & 8d & 0.8126 & 0.3112 & 0.1930\\
\hline
&\multicolumn{4}{c}{HS experimental, Ref.~\onlinecite{marezio1970}} \\
\hline
Lu  & 4c & 0.01997 & 0.     & 0.07149 \\
Fe  & 4a & 0.     & 0.     & 0.      \\
O 1 & 4c & 0.38010 & 1/2    & 0.04610 \\
O 2 & 8d & 0.81070 & 0.31210 & 0.19290\\
\hline
&\multicolumn{4}{c}{LS at 50~GPa, this work} \\
 \hline
Lu  & 4c & 0.0371 & 0.     & 0.0854 \\
Fe  & 4a & 0.     & 0.     & 0.  \\
O 1 & 4c & 0.3941 &1/2     & 0.0300 \\
O 2 & 8d & 0.8267 & 0.2987 & 0.1917\\
\end{tabular}
\end{ruledtabular}
\end{table}

\subsection{Dependence of the transition pressure upon $U$ and $J$} \label{sec:res:trans_uj}

\begin{figure}
 \begin{center}
 \includegraphics[width=0.45\textwidth]{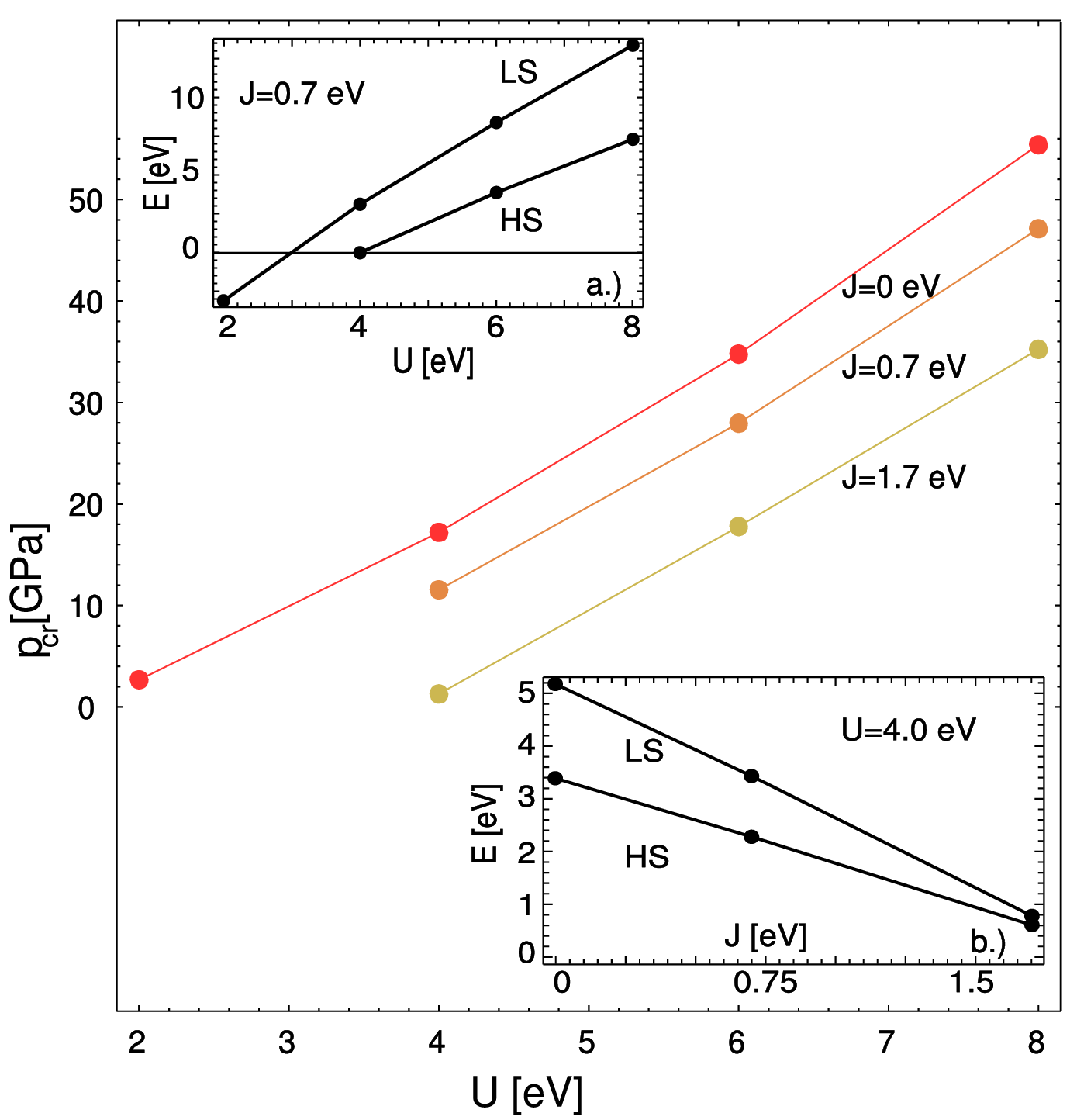}
 \caption{ \label{pic:res:uji}
The dependence of the transition pressure upon the Coulomb repulsion $U$ and the exchange parameter $J$ for the unrelaxed structures. An increase of $U$ always increases $p_{\mathrm{cr}}$ while an increase of $J$ monotonically decreases $p_{\mathrm{cr}}$. Insets: the dependence of the ground state energy on $U$ and $J$ at a fixed volume, a.) at 174.41~\AA$^3$ and b.) at 183.02~\AA$^3$.
  }
 \end{center}
\end{figure}

A number of calculations were performed with varying Hubbard parameter $U$ and exchange parameter $J$. The aim was to determine the dependence and the sensitivity of the transition pressure on these parameters. We chose the parameters in the reasonable range of [in eV] $U=\{$2, 4, 6, 8$\}$ and $J=\{$0, 0.7, 1.7$\}$. The transition pressures were calculated using the frozen ionic configuration in the cubic structure:  we were only interested in qualitative trends. The results are visualized in FIG.~\ref{pic:res:uji}. The transition pressure $p_{\mathrm{cr}}$ in the region explored can be recast as

\begin{equation} \label{eqn:trans_uj1}
p_{\mathrm{cr}}=  A  + B \, U + C\,  J
\end{equation}

with $A=-16.2$~GPa, $B=8.75$~GPa/eV and $C=- 10.7$~GPa/eV. This expression was obtained from a least squares fit to the calculated transition pressures and its reliability is $\pm1$~GPa in the range studied. The resulting coefficients $B$ and $C$ are comparable with opposite sign, which emphasizes that the main parameter of the calculation is the difference $U-J$. It shows that the formulation of \citet{dudarev1998} is a good approximation in the present case. 
In these unrelaxed calculations the increase of $U$ from 4~eV to 8~eV raises the amount of the volume collapse insignificantly from 7.53\% to 7.61\%.

The positive dependence of $p_{\mathrm{cr}}$ on $U$ can be explained as following:
Consider the simplified atomic Hamiltonian

\begin{equation} \label{eqn:trans_uj2}
\begin{split}
E_{\mathrm{ee}}&=U\sum_{i,\,j}n_i^\uparrow n_j^\downarrow+(U-J)\sum_{\sigma ,\, i>j}n_i^\sigma n_j^\sigma\\
E_{\mathrm{U}}&=E_{\mathrm{ee}}-E_{\mathrm{dc}}
\end{split}
\end{equation}

where the density matrix is diagonal ($n_{ij}^\sigma=n_i^\sigma \delta_{ij}$) and $E_{\mathrm{dc}}$ is chosen in the atomic limit (Eq. \ref{eqn:setup_1}). $E_{\rm U}$ can be recast as $\frac{U-J}{2}(N-\sum_{i,\sigma}(n_i^{\sigma})^2)$. As emphasized before,\cite{solovyev1994,cococcioni2005} this quantity cancels when occupation numbers $ n_i^{\sigma}$ are integers and is positive elsewhere (see FIG.1 of Ref.~\onlinecite{solovyev1994}). At this point, we emphasize that the effect of $J$ is thus not obvious: An increase of $J$ stabilizes the high spin state both in $E_{\rm ee}$ and in $E_{\rm dc}$, so that a clear effect on $E_{\rm tot}$ cannot be simply anticipated as emphasized before for MnO.\cite{kasinathan2007}

When going from HS to the LS hybridization effects are enhanced. This is partly due to the decrease of the volume but also due to the different spin configurations of the HS and the LS phases as they were presented in section \ref{sec:meth:dos} where the reduction of the magnetic moment of the HS phase was discussed. As a consequence, \emph{d}-orbitals are increasingly hybridized with \emph{p}-orbitals. It implies that Bloch states have a mixed  O~$p$-Fe~$d$ character: The $d$ states which should be empty in the pure ionic picture (see insets in  \ref{pic:met:hs_GPa70_pardos} and \ref{pic:met:ls_GPa130_pardos}) are more filled in the LS phase than in the HS spin phase (the occupancy is still lower than 0.5). As a consequence: $E^{\rm LS}_{\rm U} > E^{\rm HS}_{\rm U} $ (see FIG.~1 of Ref.~\onlinecite{solovyev1994}).  Thus, if $U$ is increased or $J$ is decreased, the LS phase is destabilized with respect to the HS, and thus the transition pressure increases. This is what is observed in FIG.~\ref{pic:res:uji}. This effect shows that $U$ and $J$ cannot be taken as parameters in the calculation, because the double counting expression -- though approximate -- should correct the LDA energy. This emphasizes the need for a correct determination of $U$ and $J$.
Additionally the effect of $J$ is slightly more important than the effect of $U$. This is due to the difference between expression \ref{eqn:trans_uj2}  for $E_U$ and the rotationally invariant expression that we use.\footnote{This difference is called $E_U^{\rm aniso}$ in Refs.~\onlinecite{kasinathan2007, ylvisaker2009}}

The increase of $J$ induces a charge redistribution, which further lowers the relative energy of the LS phase (see also FIG.~\ref{pic:res:uji} insets), i.e. when $J$ increases in the LS phase, electrons are transferred from the fully occupied states to the weakly occupied states. This transfer can be taken into account in the calculation of the LDA+U contribution in Eq.~\ref{eqn:trans_uj2}. We calculate this contribution for a fixed occupation at different values of $J$ and compare it with the energy shift for different values of $J$ and relaxed atomic occupancies. We find that in the second case the energy change with $J$ is three times bigger than in the first case. On the other hand, the energy change due to electron redistribution is negligible in the case of the increase of $U$. The total number of localized electrons varies when $U$ and $J$ vary. This effect is an order of magnitude smaller than the two other contributions (i.e. 1$^\mathrm{st}$ the change of $E_U$ througth an increase of $(U-J)$ with fixed occupancies and 2$^\mathrm{nd}$ the change of $E_U$ through a charge redistribution with a fixed total number of electrons).

As mentioned before, the bulk modulus is underestimated by about 20\%, which increases the deviation in volume from 2.5\% to 6\%, when pressure increases from 0 to 150~GPa (see FIG.~\ref{pic:res:lufeo3_p_V_comp}). Our analysis of $B_0(U-J)$ shows a negative correlation between $U-J$ and $B_0$ (with a slope of -30~GPa/eV), indicating that the value of $U-J$ was slightly overestimated in our structural calculations. In LuFeO$_3$ the electron-electron interaction is more efficiently screened than in FeO, where the interaction parameter was originally derived.\cite{cococcioni2005}

It is possible, that metallization is the origin of the hardening of the material at high pressure. As was observed for MnO\cite{kunes2008} and BiFeO$_3$,\cite{gavriliuk2008} the HS-LS transition can be accompanied by a insulator-metal transition.\cite{gavriliuk2008} The experimental observation could be better described by DMFT calculations.\cite{kunes2008} The increased bonding due to metallicity could make the material somewhat harder. 

\section{Conclusions } \label{sec:con}

We have carried out LDA+U calculation within the PAW framework on the rare earth perovskite LuFeO$_3$. We describe an iso-structural phase transition from a high spin phase $t_{\rm 2g}^{\downarrow 3}e_{\rm g}^{\downarrow 2}$ towards a low spin phase $t_{\rm 2g}^{\uparrow 2}t_{\rm 2g}^{\downarrow 3}$ with a volume collapse of 6.0\% (Exp: 5.5\%). Atomic positions, magnetic moments and lattice constants are computed and are in good agreement with experimental data.\cite{rozenberg2005} At high pressure, the disagreement on volume is at most 6\%. The observed reduction of the local magnetic moment on iron is $\simeq 3~\mu_B$.

We find, that the LDA\cite{iglesias2005, xing2007} is not apt to treat LuFeO$_3$. The LDA+U calculations presented here are always superior because the band gap, the phase transition pressure and the local magnetic moments could be determined more correctly.

As the computation of $U$ is not the goal of this work, we check the effect of $U$ and $J$ on the transition. We compare the filling of orbitals in the HS and LS phases and propose an interpretation for the non intuitive effect of $J$.
The determination of the parameter of the Coulombic on-site repulsion $U$ and the exchange energy $J$ appears to be essential, because the critical pressure for the spin collapse depends linearly on them. They enter the expression for the critical pressure with opposite signs but the same magnitudes [Eq.~\ref{eqn:trans_uj1}] as shown using the simplified LDA+U formalism of \citet{dudarev1998}

These calculations open the way to other complex systems such as orthoferrites where correlation effects are important both on iron and on the rare earth atom. Concerning LuFeO$_3$, some improvement of the understanding of the transition could also be brought by experimental studies of the transition as a function of temperature. Experiments on optical properties as a function of pressure as well as LDA+DMFT calculations could further clarify the nature of the transition.

\section{Acknowledgements } \label{sec:ackno}

We thank M. Torrent for the iron PAW dataset and for useful discussions. We are indebted to F. Jollet, S. Mazevet, and B. Siberchicot for useful discussions and remarks. This work was supported by the French ANR under project CORRELMAT and computations were performed at CCRT Bruyères-le-Chatel.

\bibliographystyle{apsrev}
\bibliography{/cea/dsku/trekking/hal1/home/pmc/adams/biblio/bibliography}
% biblio_v10 /cea/dsku/trekking/hal1/home/pmc/adams/biblio/bibliography

\end{document}